\documentclass[%
 reprint,
showkeys,
 amsmath,amssymb,
 aps,
]{revtex4-2}

\usepackage{graphicx}
\usepackage{siunitx}
\usepackage{dcolumn}
\usepackage{xr}
\usepackage[utf8]{inputenc}
\usepackage{bm}

\begin{document}

\preprint{APS/123-QED}

\title{Characterization of the dissolution  of water microdroplets in oil}

\author{Tamás Gerecsei $^{1,2}$}
\author{Rita Ungai-Salanki $^1$}
\author{Andras Saftics $^2$}
\author{Imre Derényi $^{1,3}$}
\author{Robert Horvath $^2$}
\author{Balint Szabo $^1$ \footnote{Corresponding autho (email: balintszabo1@gmail.com)}}

\affiliation{$^1$ Department of Biological Physics, Eötvös Loránd University, Budapest, Hungary}

\affiliation{$^2$ Centre for Energy Research, Institute of Technical Physics and Materials Science, Nanobiosensorics Laboratory, Budapest, Hungary}

\affiliation{$^3$ MTA-ELTE Statistical and Biological Physics Research Group, E\"otv\"os Lor\'and Research Network (ELKH), Budapest, Hungary}


\begin{abstract}
Water in oil emulsions have a wide range of applications from chemical technology to microfluidics, where the stability of water droplets is of paramount importance. Here using an accessible and easily reproducible experimental setup we describe and characterize the dissolution of water in oil, which renders nanoliter-sized droplets unstable, resulting in their shrinkage and disappearance in a time scale of hours. This process has applicability in creating miniature reactors for crystallization. We test multiple oils and their combinations with surfactants exhibiting widely different rates of dissolution. We derived simple analytical equations to determine the product of the diffusion coefficient and the relative saturation density of water in oil from the measured dissolution data. By measuring the moisture content of mineral and silicone oils with Karl Fischer titration before and after saturating them with water, we calculated the diffusion coefficient of water in these two oils.
\end{abstract}


\maketitle


\section{Introduction}

Water in oil (w/o) emulsions consist of a continuous oil phase with dispersed water droplets. Such arrangements are common both in industrial and laboratory settings \cite{griffiths2006miniaturising,tadros1994fundamental}. Droplet microfluidics uses w/o emulsions for containing reagents and chemical reactions such as polymerase chain reaction (PCR) and others for DNA/RNA sequencing \cite{weitz2017perspective,ding2017single}. 
The droplet-based approach has several advantages, mainly that the volume of the droplets not only matches the desirable size range for single-cell manipulations, but it also minimizes the amount of reagents needed. Such setups have been successfully commercialized and they proved to be a revolutionary tool in single-cell analysis \cite{shembekar2016droplet} enabling the development of lab-on-a-chip devices  \cite{haeberle2007microfluidic} that are capable of integrating entire bioassay workflows on handheld microfluidic chips.
Specifically, many systems use sessile aqueous droplets printed by microfluidic robots \cite{zhu2015printing}.
These applications center around the printing of aqueous droplets under oils for protein engineering \cite{zhu2014nanoliter},  genome amplification by PCR \cite{white2011high} and transpriptomics \cite{dalerba2011single}.

Additive manufacturing, i.e., three dimensional printing of various materials has been gaining tremendous momentum in recent years. Regenerative medicine applying tissue engineering can greatly benefit from this revolution through technologies such as bioprinting of tissues or organs \cite{murphy20143d}. To achieve single cell resolution in bioprinting, several droplet printing solutions has been proposed with the ultimate goal of trapping single-cells inside subnanoliter sized droplets \cite{francz2020subnanoliter,zhu2013sequential}. The time scale of keeping single cells in tiny droplets can vary from a few seconds to several hours or days in long term assays.

Once created, the volume of the droplets needs to stay constant for the entire duration of the workflow without merging or breaking up. If the volume were to change during an assay or experiment, the concentration of reagents in the droplet would be altered, thus, one of the fundamental functions of the carrier oil
(volume conservation by impeding evaporation) would diminish. 

Merging is an important issue in a dense emulsion, when the overall volume of the water is comparable to that of the oil \cite{mazutis2013single,shah2008designer}. In these systems there is a physical contact between the interfaces of the droplets that needs to be stabilized by the addition of surfactants \cite{baret2012surfactants}. These amphipathic substances associate on the oil/water interface and prevent the coalescence of droplets. 

Without any protective oil layer, sessile droplets rapidly evaporate into the atmosphere. This process has been investigated in depth both theoretically \cite{picknett1977evaporation} and experimentally \cite{soolaman2005water,erbil2002drop}. In general, two modes of evaporation from a solid surface have been identified: in the so-called 'pinning' mode the contact area of the droplet formed with the solid surface is constant, while the contact angle decreases, whereas in the 'shrinking' mode, the contact area shrinks, while the contact angle remains unchanged. A mixed mode where both quantities are simultaneously diminishing has also been described and experimentally observed \cite{erbil2002drop}.  

To stabilize the volume of droplets, cover oils can be used. Although water and oil molecules are immiscible due to their polar and apolar nature, a small degree of water solubilization does occur in oil. This phenomenon usually can be neglected for macroscopic droplets, however, micrometer scale droplets exhibit a gradual loss of volume over time in an unsaturated oil. Mass transfer of a liquid droplet into another miscible liquid has been modeled theoretically by Epstein and Plesset \cite{epstein1950stability} inspiring several experimental studies \cite{su2013mass,duncan2006microdroplet}. 
Schmitt et al. considered the mass transfer of water droplets in an immiscible oil environment with added emulsifier (Span80) \cite{schmitt2017spontaneous}. They found that the droplet surface exhibited spontaneous microsctructure formation decreasing the volume and distorting the shape of the initially spherical droplets. In a study by Rodríguez-Ruiz et al.\ nanoliter sized droplets of $NaCl$ solution under silicone oil were found to dissolve in a mixed mode \cite{rodriguez2013monitoring}.

While in most microfluidic devices the stability of the emulsion is needed, some applications require shrinking droplets, the volume of which decreases in a controlled manner. In such systems a high level of control over the dissolution rate of droplets is desirable. The theory behind the dissolution of microdroplets in various immiscible liquids has been established over the years \cite{zhang2015mixed}. When droplets under oil contain an aqueous solution, the loss of water through dissolution leads to increasing solute concentration and potentially to crystallization \cite{velazquez1970studies,grossier2010ultra}, although the concentration of the solute is known to modify the rate of shrinking \cite{bitterfield2016activity, utoft2018manipulating}. Therefore, dissolving microdroplets show great promise in their applicability as miniature reactors for solvent-diffusion-driven crystallization to produce protein crystals or solid drug dosages \cite{pal2020mathematical,espitalier1997spherical}.  Furthermore, diffusion of water into the surrounding oil phase on a microfluidic chip has been used to create hydrogel microparticles with a homogeneous and modifiable size distribution \cite{pittermannova2016microfluidic}.

Regardless of the application, any oil used in microfluidics needs to comply with some basic requirements such as chemical inertness, biocompatibility, good optical qualities (transparency) and the existence of compatible surfactants. The most popular oils that fulfill the above requirements are mineral, silicone, and fluorinated oils \cite{baret2012surfactants}.
 
In the present paper we investigate the long-term behavior of water droplets in sparse w/o emulsions. Emulsions in various oils were generated using a simple rotating fluid-based method. Subsequently, the  droplets were imaged by time-lapse microscopy and a modified contact angle measurement setup to investigate the evolution of their shape descriptors. 
We provide a straightforward and easily reproducible workflow for the experimental characterization of sessile droplet dissolution. We show that when the droplets are hemispherical, the product of the diffusion coefficient and the saturation density can be analytically calculated.

\section{Materials and Methods}

\subsection{Droplet generation}

\begin{figure}
    \centering
    \includegraphics[width=0.5\textwidth]{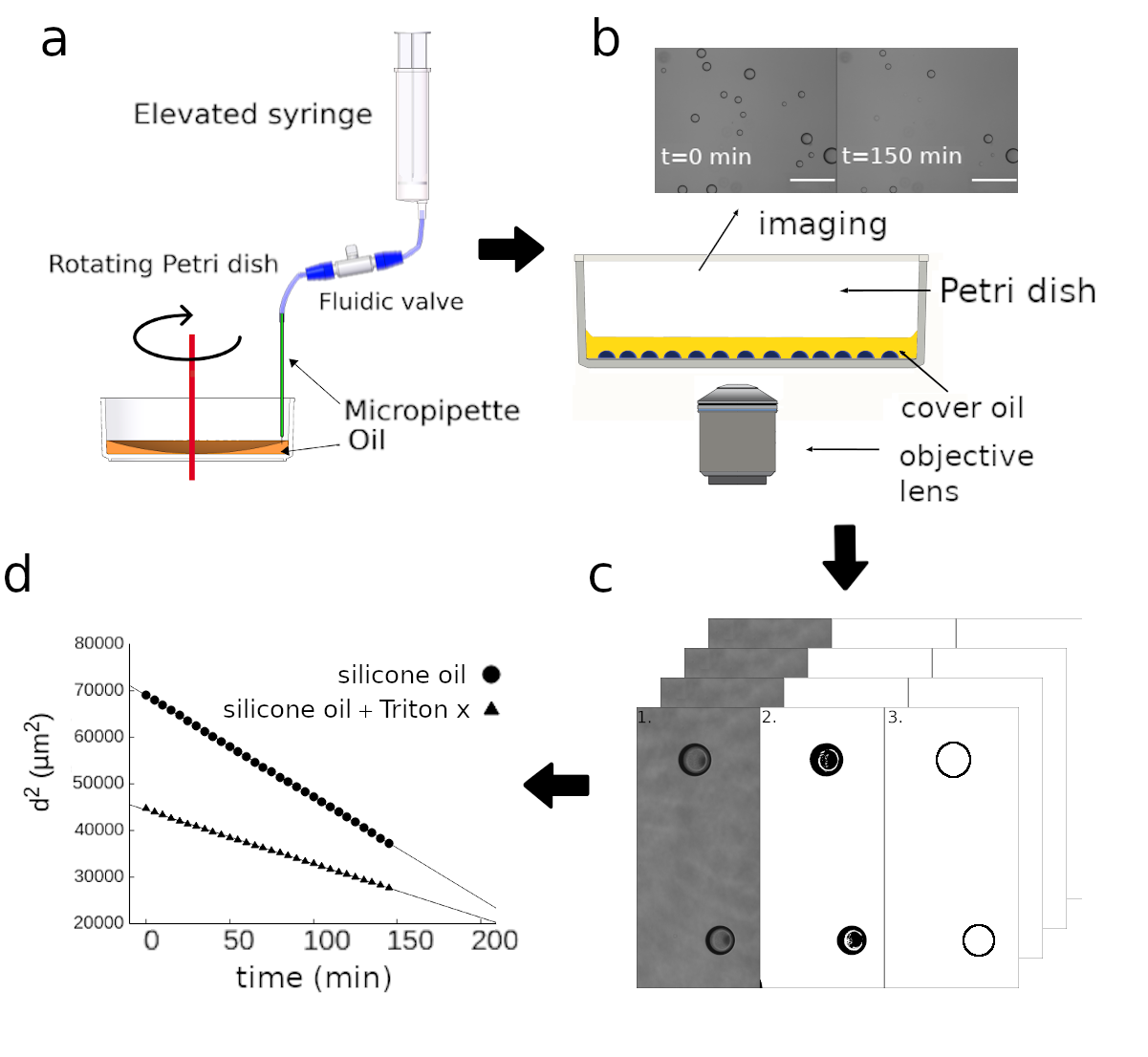}
    \caption{The workflow of droplet diameter measurements. \textbf{a:} The droplets are generated by a water filled micropipette immersed in a rotating Petri dish containing oil. \textbf{b:} The Petri dish with the w/o emulsion is placed on an inverted microscope and images are taken in time-lapse mode. Scale bars indicate $100\,\si{\micro\metre}$. \textbf{c:} Image stacks are segmented to detect and track the droplets. \textbf{d:} Diameter of droplets is automatically measured to plot the curves of droplet dissolution.}
    \label{fig:workflow}
\end{figure}

Water in oil emulsions were generated using a rotating fluid-based method (similar to \cite{umbanhowar2000monodisperse}) illustrated in Fig.\ref{fig:workflow}.a. A standard plastic, hydrophobic Petri dish (Greiner) with a diameter of $35\,\si{\milli \metre}$ was placed onto a rotating platform, then filled with $1\,\si{\milli \litre}$ oil using a handheld pipette. The following oils were investigated: mineral oil (Sigma M8410, density: $0.85\, g/ml$ ), mineral oil mixed with $0.5\,(\mathrm{vol}\si{\percent})$ Span80 (Fluka Analytical), silicone oil (AR 20, density: $1.01\, g/ml$ ) and silicone oil mixed with $0.2\,(\mathrm{vol}\si{\percent})$ Triton X-100.
All materials were purchased from Sigma-Aldrich unless otherwise stated. The custom made platform was rotated by a PC fan (Coolink). Frequency of spinning was monitored by the tachometric output of the fan. We applied $360$ rotations per minute (rpm) corresponding to a speed of $~ 37\,\si[per-mode=symbol]{\centi \metre \per \second }$ at the tip of the micropipette.

Once the dish was spinning, a glass micropipette (inner diameter $d=50\,\si{\micro \metre}$) was immersed into the oil from above to a depth of max $1\,\si{\milli\metre}$ and a distance from the axis of rotation of $1\,\si{\centi\metre}$. The other end of the micropipette was connected to an elevated plastic syringe through a PTFE tube with an inner diameter of 1 mm, interrupted by a normally closed valve. The entire fluidic system was filled with deionized water (Seralpur AP30, Seral). After immersing the tip of the micropipette in the oil, the valve was opened for one minute. During this time, a flow commenced through the micropipette due to the applied hydrostatic pressure of $2500\,\si{\pascal}$. As the water entered the rotating oil from the micropipette, it was separated into nanoliter scale droplets by the shear force acting on its surface. After one minute, the micropipette was removed from the oil, the rotation was stopped, and the Petri dish with the droplets inside was placed onto the microscope for time-lapse analysis (Fig.\ref{fig:workflow}.b).
The oil layer in the Petri dish was $~1\,\si{\milli\metre}$ thick with its top surface exposed to the room temperature ($~23\,\si{\degreeCelsius}$),  unsaturated air. Using such a setup,  water molecules can diffuse from the oil to the air. This condition ensures the permanent low concentration of dissolved water in oil at this surface. Using dry air or nitrogen above the dish can further improve the precision of the experiment.

\subsection{Time-lapse analysis}

Time-lapse recordings of the droplets were executed using an inverted microscope (Zeiss Axio Observer) equipped with a 10x EC Plan Neofluar objective, CMOS camera (Andor Zyla 5.5) and motorized stage (Marzhauser). The Petri dish with the w/o emulsion inside was placed onto the sample holder. Subsequently, fields of view (FoVs) with surface attached drops were identified and time-lapse imaging was programmed using the CellSorter software. Each FoV was recorded in brightfield mode every $5$ minutes for a period of time that varied by experiment. For each FoV we used an auto-focusing algorithm to follow the equator of the droplets moving out of the focal plane as they shrank. A z-stack of three images was captured with a distance of $5\,\si{\micro\metre}$ between the z-planes. Software chose the sharpest z-plane.  The recorded images were stored for later analysis.

To determine the change of shape of the droplets from a side view, a custom developed contact angle measurement setup (Plósz Mérnökiroda Ltd.) was used. A glass cuvette was filled with oil and a separated Petri dish bottom was placed into it. This was necessary in order to have the drops attach to the same surface as in the experiments with the inverted microscope described above. Water droplets were generated manually under the oil using a glass micropipette with an inner diameter of $50\,\si{\micro\metre}$. Once a droplet with a size in the target range ($\approx 500\,\si{\micro\metre}$) sedimented onto the surface, the magnifying optics was focused and the time-lapse imaging was started. Images were recorded every $15$ minutes for a period of several days until the examined droplet completely disappeared.

\subsection{Image analysis}

Time-lapse images were analyzed with the ImageJ software \cite{schindelin2012fiji}. Using an inverted microscope we could measure the diameter of the droplets with a shape of a spherical cap from a bottom view. Every field of view was treated as an image sequence in which the changing diameter of the droplets needs to be determined on each frame. First, the region of interest (RoI) in a given FoV was cut out (Fig.\ref{fig:workflow}.c.1). Afterwards, a threshold was applied to each image using the Triangle algorithm \cite{zack1977automatic}, which separated the droplets from the background resulting in a binary image (Fig.\ref{fig:workflow}.c.2). Subsequently, the built-in particle analyzer algorithm of ImageJ was applied to identify the droplets (Fig.\ref{fig:workflow}.c.3) and measure their diameter. The recorded values of diameter were then plotted against time and further analyzed (Fig.\ref{fig:workflow}.d).

In case of the contact angle measurement, we used the DropSnake plugin \cite{stalder2006snake} to analyze the side view images of the droplets (see Fig.~S2). Contact angles were extracted every $2.5\,\si{hours}$. Around $8$ points along the outline of the droplets were manually highlighted and then the edge was automatically fit by the algorithm. The contact diameter on the solid surface, the droplet diameter, and the left and right contact angles were calculated. The average of these two angles was accepted as the final contact angle value. 

\subsection{Moisture determination of oils}

We determined the moisture content of the mineral and silicone oils both in their factory qualities and after saturation them with water. We saturated 20 ml of the oils by mixing them with deionized water with a magnetic stirrer (Biosan MSH 300) in a closed 50 ml upside down centrifuge tube at room temperature for more than 24 hours. The water to oil ratio was 50\% in the tube. We used the maximum 1250 RPM speed of the stirrer to achieve an oil droplet size smaller than $1\si{mm}$. We centrifuged the saturated oils 2 times at $300\, G$ for $10$ min to remove the remaining water droplets.

Moisture content of the oils was measured in a Kyoto MKS-500 Karl Fischer moisture titrator.

\section{Results}

\subsection{Time-lapse analysis of droplets}

\begin{figure}[htb]
    \centering
    \includegraphics[width=0.5\textwidth]{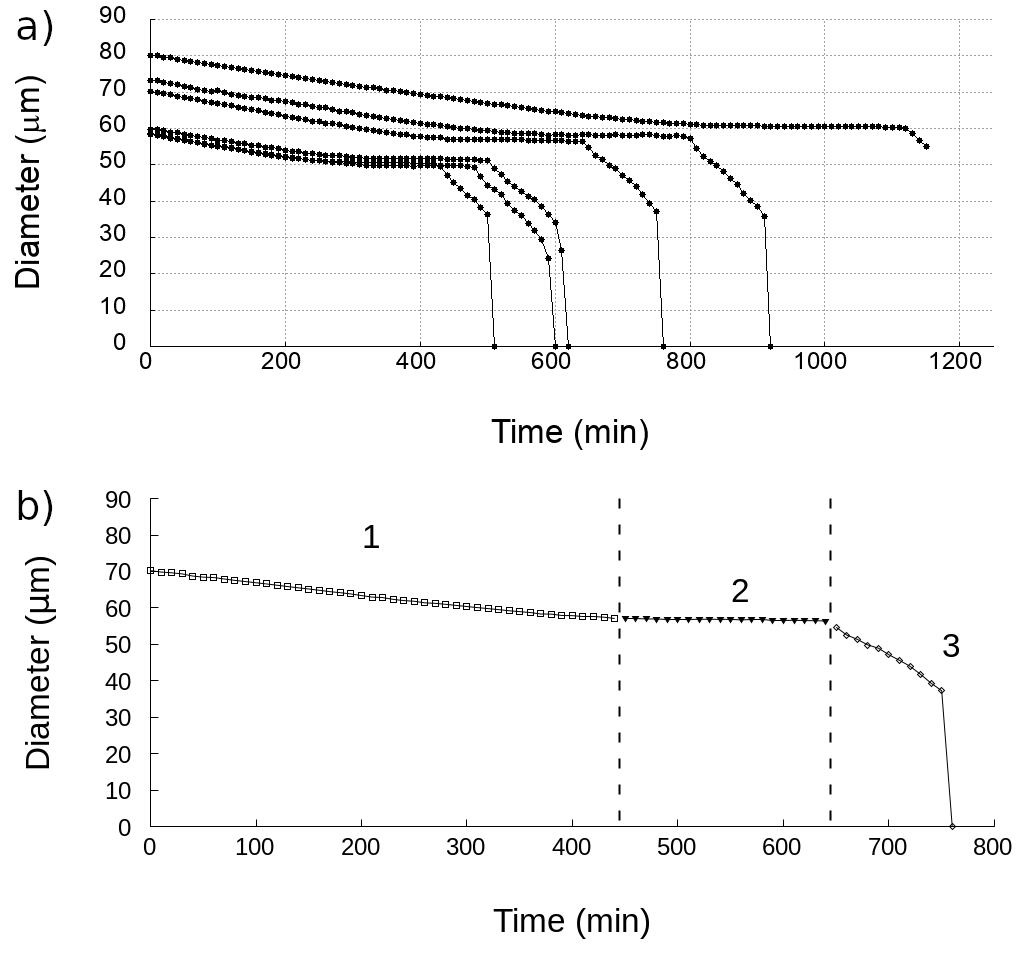}
    \caption{a) Data showing the change of the diameter of water droplets under mineral oil from a bottom view as observed by the objective lens of an inverted microscope. Different curves correspond to different droplets. b) The three distinctive phases of a representative $d(t)$ curve. }
    \label{fig:curves}
    \end{figure}

\begin{figure}
    \centering
    \includegraphics[width=0.5\textwidth]{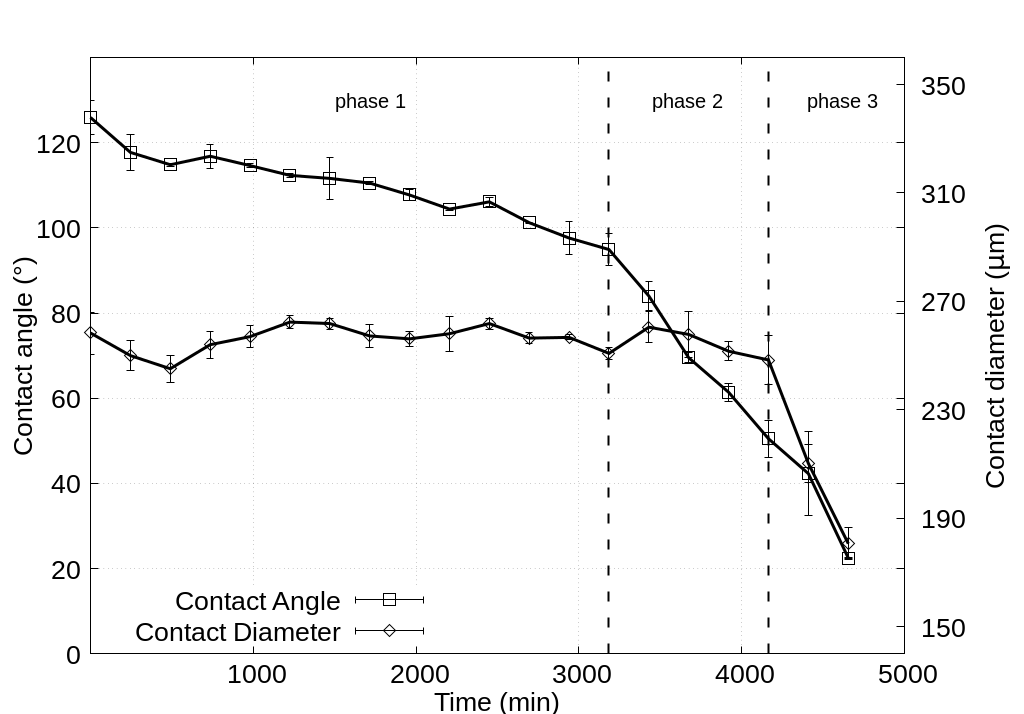}
    \caption{The time evolution of shape descriptors of a droplet imaged from a side view. Vertical lines indicate the boundary between the three different phases. See also supplementary video 2.}
    \label{fig:double}
\end{figure}

Generation of droplets was executed as described above. In a typical run $2000-5000$ droplets were generated with a relatively low width of size distribution (Fig.~S1). Then the hydrophobic Petri dish containing the w/o emulsion of thousands of droplets was placed onto an inverted microscope and let to sediment for $5\,\si{\minute}$. Time-lapse images were recorded to simultaneously observe multiple, similarly sized droplets under the same conditions. 
The change of the diameter of a number of such water droplets under mineral oil can be seen in Fig.~\ref{fig:curves}a and Supplementary video~1. A perfect immiscibility between water and oil would dictate a constant size, however, time-lapse recordings show a gradual dissolution of the droplets. As indicated by a representative curve, the decrease in the diameter $d(t)$ of the droplets from a bottom view occurs in three distinct stages (labeled with numbers 1-3) beginning with a linear decrease, followed by a stagnating phase, and ending with a rapider decrease (Fig.~\ref{fig:curves}b). In order to better understand this phenomenon, we executed an another type of experiment: a water droplet placed under mineral oil was imaged from side view using a contact angle measurement setup. In both experiments the droplets were deposited on a hydrophobic plastic Petri dish surface, exhibiting an initial contact angle higher than $90^{\circ}$. The change of the contact angle and the contact diameter imaged from a side view can be seen in Fig.~\ref{fig:double}. It is apparent that the contact area between the droplet and the surface does not change in phases 1-2, but it quickly shrinks in the final phase 3 of droplet dissolution.


\begin{figure}
    \centering
    \includegraphics[width=0.5\textwidth]{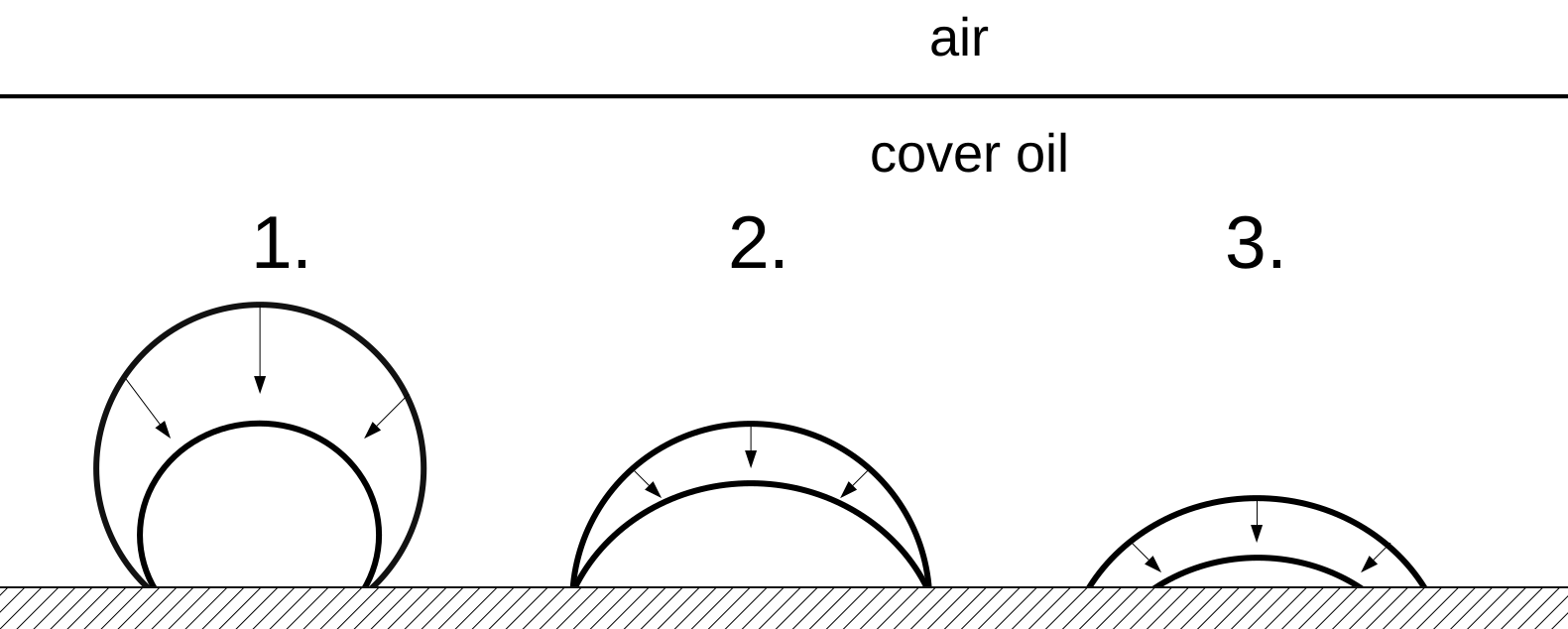}
    \caption{3 phase model of sessile droplet dissolution under a cover oil on a hydrophobic surface. Phase 1: The contact angle decreases together with the droplet diameter, while the contact area remains unchanged. Phase 2: After the contact angle reaches $90^{\circ}$, it keeps decreasing together with the volume of the droplet but with a constant diameter of its vertical projection being equal to the contact diameter. Phases 1 and 2 correspond to a pinning mode dissolution. Phase 3: The contact diameter and the contact angle decrease at the same time until complete disappearance of the droplet corresponding to a mixed mode dissolution.}
    \label{fig:drawing}
\end{figure}

According to these results a model can be proposed for the dissolution of the droplets and the change of their shape descriptors. During the entire process, the droplet shape can be described as a spherical cap sitting on a flat surface. In phase 1 the radius of the droplet decreases gradually due to the loss of droplet volume. In this phase the contact angle decreases, as well. Since the contact area does not change (Fig.~\ref{fig:double}.), this phase corresponds to the pinning mode of evaporation (Fig.~\ref{fig:drawing}.1). Phase 2 begins when the contact angle reaches $90^{\circ}$. At this point the droplet is a hemisphere. Then the contact angle and volume of the droplet keep decreasing but with a constant diameter of its vertical projection being equal to the contact diameter (Fig.~\ref{fig:drawing}.2). In phase 3, the contact surface area rapidly shrinks together with the contact angle until complete dissolution. Thus this phase corresponds to a mixed mode (Fig.~\ref{fig:drawing}.3), whereas phases 1 and 2 correspond to the pinning mode of dissolution.

To investigate the generality of the phenomenon and identify any surfactant-induced effect, droplets under different oil mixtures were imaged in time-lapse mode on an inverted microscope. 

\begin{figure}
    \centering
    \includegraphics[width=0.5\textwidth]{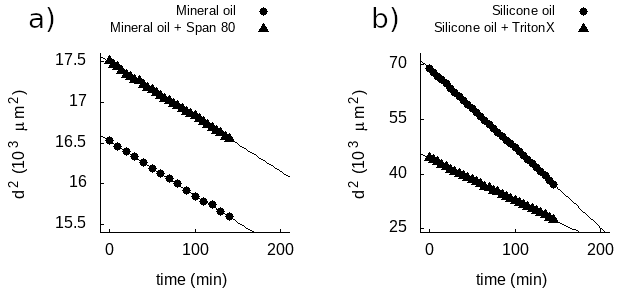}
    \caption{Square of the droplet diameter $d^2(t)$ as a function of time in phase 1 measured on an inverted microscope. a) $d^2(t)$ curves for representative droplets in mineral oil and mineral oil mixed with surfactant (Span80). b) $d^2(t)$ curves for representative droplets in silicone oil and silicone oil mixed with surfactant (Triton X-100). Data series were fit with a line. Relative standard error of the slope was $<1\%$ in all cases.}
    \label{fig:d2t2}
\end{figure}

According to the empirical theory proposed by Picknett \cite{picknett1977evaporation} and Erbil \cite{erbil2002drop} for spherical droplets, the two third power of the volume of the droplets ($V^{2/3}$) should show a linear decrease in time. Thus, the square of the droplet diameter ($d^2$) is also expected to show a linear decrease in phase 1. This behaviour has indeed been reproduced by our time-lapse measurements for all four oil mixtures we studied in depth as Fig.~\ref{fig:d2t2} demonstrates. Droplets on a glass surface showed identical behaviour (Supplementary video 3).
In phase 1 both the volume and the contact angle of the droplets can be calculated from the measurements carried out on the inverted microscope (Fig.~S3-5). However, the radius of the spherical cap in phase 2 and 3 cannot be easily measured using this setup. We could follow this radius until the end of the process using the contact angle measurement device. Interestingly, this quantity has a minimum when the shape of the droplet is a hemisphere (Fig.~S6).

It is apparent, that the droplet size decreased significantly faster in silicone oil than in mineral oil. 
The presence of surfactants did not impede dissolution. Amphiphilic molecules present in the oils are expected to form a monolayer on the oil-water interface but do not block water molecules to diffuse out of the droplet. However, according to our results their presence appears to affect the dissolution rate of the droplets providing an opportunity to fine-tune the process. 

We observed spontaneous microstructure formation on the surface of the water droplets in oils containing surfactant (Fig.~S.7). During the dissolution process the droplet surface became cloudy and dark, exhibiting a rough morphology resembling the observations by Schmitt et al. \cite{schmitt2017spontaneous}.


\subsection{Calculation of the diffusion coefficient and saturation density}

The measurements presented here allow us to calculate some of the physical parameters characterizing the diffusion and solubility of water in oil. Water diffusion can be described by the diffusion equation
\begin{equation}
    \frac{\partial\rho(\mathbf{x},t)}{\partial t}=D\Delta\rho(\mathbf{x},t),
\end{equation}
where $\rho(\mathbf{x},t)$ denotes the mass density of water as a function of space ($\mathbf{x}$) and time ($t$); and $D$ is its diffusion coefficient. Assuming stationarity (which is a good approximation if the oil-water interface moves much slower than the speed of the diffusion over the length-scale of the droplet) the diffusion equation simplifies to Laplace's equation:
\begin{equation}
     \Delta\rho(\mathbf{x})=0.
\end{equation}
For a spherical droplet or, equivalently, when a hemispherical droplet is sitting on the side of a half-space (as is the case at the end of phase 1 in our experiments) the most suitable choice for the coordinate system is the spherical one. Denoting the radial coordinate by $r$, Laplace's equation can be expressed as
\begin{equation}
    \frac{1}{r^2}\partial_r\left(r^2\partial_r \rho(r)\right)=0.
\end{equation}
Its solution that satisfies the boundary conditions 
$\rho(\infty)=\rho_\infty$ and $\rho(R)=\rho_\mathrm{s}$,
where $R$ is the radius of the droplet and $\rho_\mathrm{s}$ is the saturation density of water is
\begin{equation}
    \rho(r)=\frac{R(\rho_\mathrm{s}-\rho_\infty)}{r}+\rho_\infty.
    \label{eq:solution}
\end{equation}
The mass flux density from Fick's first law can be determined as
\begin{equation}
    j(r)=-D\partial_r\rho(r)=\frac{DR(\rho_\mathrm{s}-\rho_\infty)}{r^2},
\end{equation}
which, at the droplet boundary becomes
\begin{equation}
    j(R)=\frac{D(\rho_\mathrm{s}-\rho_\infty)}{R}.
\label{eq:flux}
\end{equation}

Effect of capillary pressure on the saturation density can be neglected as the droplet size is orders of magnitude larger than the size of water molecules \cite{webster1998stabilization}. (For details see section 5 in the Supplementary material.) This mass flux density can be directly determined from the time-lapse measurements. Close to the time point when the shape of the droplet is a hemisphere (which is equivalent to a sphere in the full space), the mass flux density is:
\begin{equation}
    j(R)=
    -\rho_0\frac{1}{A}\frac{dV}{dt}=
    -\frac{\rho_0}{2\pi R^2}\cdot2\pi R^2\frac{dR}{dt}=
    -\rho_0\frac{dR}{dt}
    \label{eq:Jm}
\end{equation}
where $\rho_0\approx10^3\,\si[per-mode=symbol]{\kilogram\per\metre^3}$ is the density of bulk water, and $A$ and $V$ denote the area and volume of the droplet of radius $R$, respectively.
Plugging this expression into Eq.~(\ref{eq:flux}) results in
\begin{equation}
    -2D\frac{\rho_\mathrm{s}-\rho_\infty}{\rho_0}=
    2R\frac{dR}{dt}=
    \frac{dR^2}{dt},
\end{equation}
which shows, in agreement with Fig.~\ref{fig:d2t2}, that $R^2$ decreases linearly with time at a rate of
\begin{equation}
    2D\frac{\rho_\mathrm{s}-\rho_\infty}{\rho_0}.
\label{eq:measured_quantity}
\end{equation}

\begin{figure}
    \centering
    \includegraphics[width=0.5\textwidth]{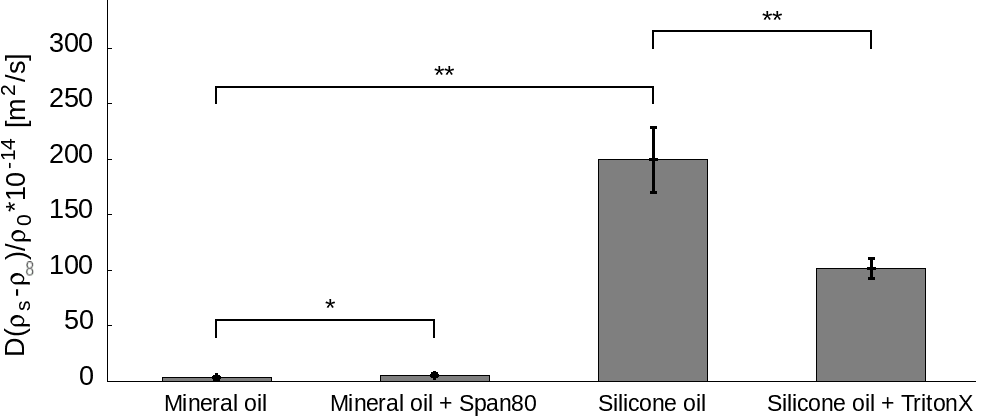}
    \caption{Values of $D(\rho_\mathrm{s}-\rho_\infty)/\rho_0$ of the four different oil mixtures we investigated in depth. For each material the data shown corresponds to a linear fit of five different droplets under the same conditions. Errors were calculated as the standard deviation of the five slopes. Significance levels: $**:p<0.01\,\%$, $*:p<0.05\,\%$. Values of $D(\rho_\mathrm{s}-\rho_\infty)/\rho_0$ are the following. Mineral oil: $(3.31\pm1.28)\times 10^{-14}\,\si[per-mode=symbol]{\metre^2\per\second}$; mineral oil with $0.5\,(\mathrm{vol}\%)$ Span80: $(5.57\pm1.41)\times 10^{-14}\,\si[per-mode=symbol]{\metre^2\per\second}$; silicone oil: $(199.57\pm29.14)\times 10^{-14}\,\si[per-mode=symbol]{\metre^2\per\second}$; silicone oil with $0.2\,(\mathrm{vol}\%)$ Triton X-100: $(101.75\pm9.36)\times 10^{-14}\,\si[per-mode=symbol]{\metre^2\per\second}$.}
    \label{fig:slopes}
\end{figure}

Measuring this rate provides the values of $D(\rho_\mathrm{s}-\rho_\infty)/\rho_0$ for the four different cover oils investigated (see Table I). This quantity characterizes the stability of the water droplets through two fundamental physical parameters: the diffusion coefficient and the saturation density of water in the specific oil. The results presented in Fig.~\ref{fig:d2t2} and summarized in Fig.~\ref{fig:slopes} show that different oils can exhibit significantly different dissolution characteristics. Interestingly, the presence of a surfactant can change droplet stability in contrasting ways. In case of mineral oil, the rate of dissolution was increased by the presence of the surfactant Span80, while for silicon oil, it was decreased by a factor of $2$ by the surfactant Triton X-100. \\
\begin{table}[h]
\centering
\begin{tabular}{ | p{2cm} | p{2cm}| p{2cm}| p{2cm} | } 
\hline
Mineral oil& Mineral oil + Span80 & Silicone oil & Silicone oil + Triton X-100 \\ 
\hline
$3.31\pm1.28$ & $5.57\pm1.41$ & $199.57\pm29.14$ & $101.75\pm9.36$\\ 
\hline
\end{tabular}
\caption{\label{tab:rho*satconc} Values of $D(\rho_\mathrm{s}-\rho_\infty)/\rho_0$ in units of $10^{-14}\,\si[per-mode=symbol]{\metre^2\per\second}$ for the four different oils.}
\end{table}
\begin{table}[h]
\centering
\begin{tabular}{ | p{2cm} | p{2cm}| p{2cm}| p{2cm} | } 
\hline
Mineral oil& Saturated mineral oil & Silicone oil & Saturated silicone oil \\ 
\hline
$341$ & $480$ & $609$ & $2902$\\ 
\hline
\end{tabular}
\caption{\label{KarlFischer} Values of the moisture content of oils in ppm as measured by Karl Fischer titration.}
\end{table}

We measured the moisture content of mineral and silicone oils with Karl Fischer titration before and after saturating them with water as shown in Table II. We calculated the diffusion coefficient of water in these two oils resulting in $2.8$ and $8.6$ $\times 10^{-10}\,\si[per-mode=symbol]{\metre^2\per\second}$ for the mineral and silicone oil, respectively.

\section{Discussion}
Testing the two most widely used oils, silicone and mineral oil, 
we observed that nanoliter-sized water droplets dispersed in oil gradually shrank and disappeared in a few hours. Application of surfactants did not have a qualitative impact on the phenomenon, however they have the potential to modify the droplet dissolution rate. We could monitor the volume and shape (contact angle) of water droplets under oil on a solid surface  using a low-cost, easily reproducible setup. Our results fit well into the previously established theoretical framework. The rate of droplet shrinking is determined by the dissolution and diffusion of water in oil allowing to calculate the product of the diffusion coefficient and the saturation density of water in oil. Data on the diffusion coefficient or saturation density of water in oils are scarce in the literature.  Although vegetable oils can absorb $\sim1000$~ppm water  \cite{hilder1968solubility}, saturation density in transformer oils (special mineral oils with low moisture content) was reported to be only $\sim70$~ppm  at room temperature ($25\,\si{\degreeCelsius}$) \cite{du2001moisture}. Diffusion coefficient of water in paraffin oil and in groundnut oil was found to be $8.5\times 10^{-10}\,\si[per-mode=symbol]{\metre^2\per\second}$ and $2.5\times 10^{-10}\,\si[per-mode=symbol]{\metre^2\per\second}$ at room temperature, respectively \cite{hilder1971diffusivity}. Our results lie in the expected range confirmed by the values of the diffusion coefficient of water in mineral ($2.8\times 10^{-10}\,\si[per-mode=symbol]{\metre^2\per\second}$) and silicone ($8.6\times 10^{-10}\,\si[per-mode=symbol]{\metre^2\per\second}$) oils. These were calculated on the basis of our droplet dissolution measurements and the moisture content of the oils before and after saturation determined by Karl Fischer titration. 

We offer a simple experimental method along with analytical calculations to determine the diffusion coefficient or saturation density of immiscible liquids using only a small volume of reagents, when either of these quantities is already known. The saturation density can be measured using a standard Karl-Fishcer moisture titrator device to determine the diffusion coefficient. 
Our results can be exploited in droplet microfluidics, nanoliter-to-picoliter-scale droplet printing, and to engineer well controlled, concentration-dependent reactions in tiny volumes, e.g., for solvent-diffusion-driven crystallization.

\begin{acknowledgments}
We are grateful to Krisztián Gál (greenlab.hu) and Prof Gyula Záray for completing the Karl Fischer titration. This work was supported by the Hungarian National Research, Development and Innovation Office (grant numbers: PD 124559 for R. U. S., KH-17, KKP 129936 and ERC-HU for R. H.), and the “Lendület” Program of the Hungarian Academy of Sciences for R. H.
\end{acknowledgments}

\appendix

\bibliography{ms}


\end{document}


\renewcommand{\figurename}{Supplementary Figure}


 
\title{Supplementary material to: Characterization of the dissolution  of water microdroplets in oil}

\author[1,2]{Tamás Gerecsei}
\author[1]{Rita Ungai-Salanki}
\author[2]{Andras Saftics}
\author[1,3]{Imre Derényi}
\author[2]{Robert Horvath}
\author[1]{Balint Szabo \footnote{Corresponding author (Email: balintszabo1@gmail.com)}}

\affil[1]{Department of Biological Physics, Eötvös Loránd University, Budapest, Hungary}

\affil[2]{Centre for Energy Research, Institute of Technical Physics and Materials Science, Nanobiosensorics Laboratory, Budapest, Hungary}

\affil[3]{MTA-ELTE Statistical and Biological Physics Research Group, Budapest, Hungary}

\maketitle




\section{Droplet generation} 

\begin{figure}[H]
    \centering
    \includegraphics[width=8.6cm]{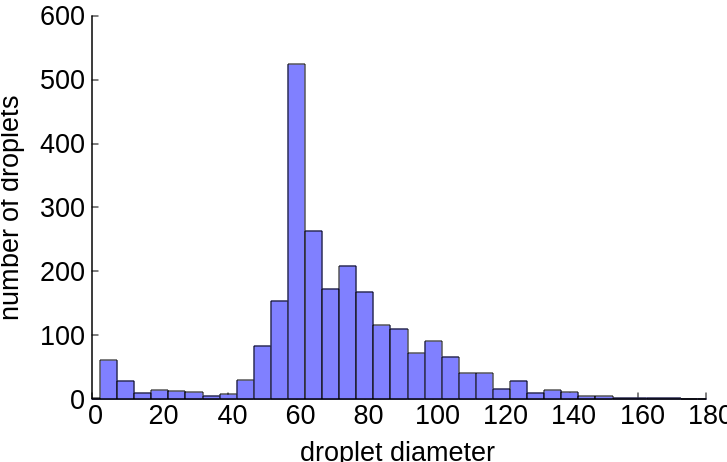}
    \caption{Distribution of droplet diameter in a typical droplet generation run. The number of droplets generated and represented here is $2397$. In a typical run $2000-5000$ droplets were created. The droplet generation method based on rotating fluid was chosen for its ability to create a large number of droplets directly in the Petri dish. 
}
    \label{fig:my_label}
\end{figure}

\section{Side view imaging}

\begin{figure}[H]
    \centering
    \includegraphics[width=8.6cm]{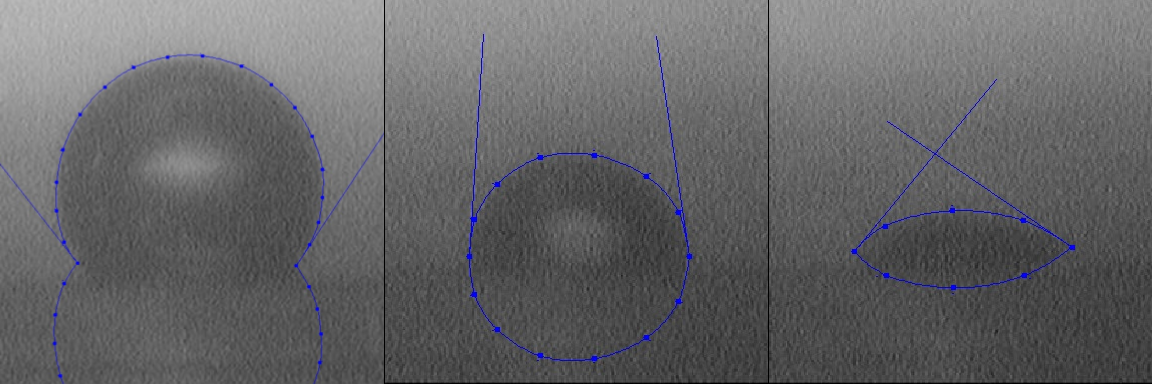}
    \caption{Evaluation of contact angle measurements in three different images. The blue points correspond to manually placed markers which were then used by the DropSnake plugin of ImageJ to fit a spherical cap on the droplets. The shape descriptors were automatically calculated and saved.}
    \label{fig:fillapse}
\end{figure}

\begin{figure}[H]
    \centering
    \includegraphics[width=8.6cm]{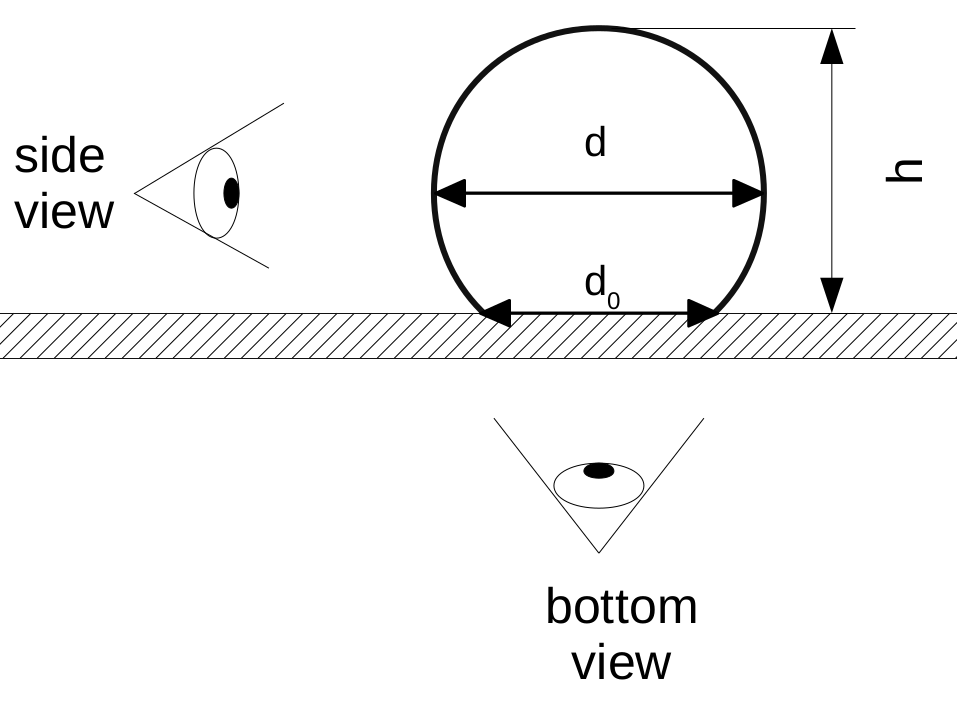}
    \caption{Geometry of a sessile droplet on a hydrophobic surface. $d$ is the diameter of the spherical droplet at its equator, while $d_0$ is the contact diameter.}
    \label{fig:droplet_views}
\end{figure}

The geometry of a sessile droplet can be described by a spherical cap. On a hydrophobic surface the contact angle is larger than $90^{\circ}$ in equilibrium. Thus the equator of the droplet with a diameter of $d$ is above the solid surface. This was the case in the first phase of dissolution in our experiments.  The diameter of the contact area with the solid surface is $d_0$. These two parameters determine the overall volume of the droplet:
\begin{equation}
    V_\mathrm{droplet}=\frac{1}{6}\pi h\
    \left[3\left(\frac{d_0}{2}\right)^2+h^2\right]
    \label{eq:vdrop}
\end{equation}{}
where the height of the droplet:
\begin{equation}
    h=\frac{d}{2}+\sqrt{\left(\frac{d}{2}\right)^2-\left(\frac{d_0}{2}\right)^2}
\end{equation}{}

As long as $d>d_0$, i.e., in phase 1 of the droplet dissolution (Fig.3.1.), $d$ can be determined on an inverted microscope by observing the droplet from a bottom view.  $d_0$ can be measured as the constant diameter of the droplet from a bottom view during the second phase of dissolution (Fig.3.2.). The contact angle ($\theta$) of the droplet can also be calculated from these two diameters:
\begin{equation}
    \theta=180^{\circ}-\arcsin\left(\frac{d_0}{d}\right).
    \label{eq:supd}
\end{equation}{}

\begin{figure}[H]
    \centering
    \includegraphics[width=8.6cm]{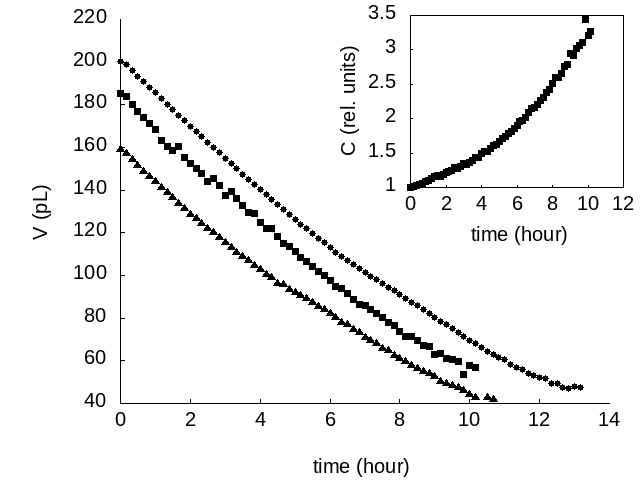}
    \caption{Volume of three representative droplets under mineral oil in phase 1 as measured on an inverted microscope, and calculated from Suppl. eq.\ref{eq:vdrop}. The inset shows the change in solute concentration occurring in a droplet as calculated from one of the volume curves. Assuming that the diffusion of the solute into the oil phase can be neglected, the solution gets progressively denser. This phenomenon can be utilized to "scan" over a concentration range, potentially leading to crystallization.}
    \label{fig:vt}
\end{figure}{}

\begin{figure}[H]
    \centering
    \includegraphics[width=8.6cm]{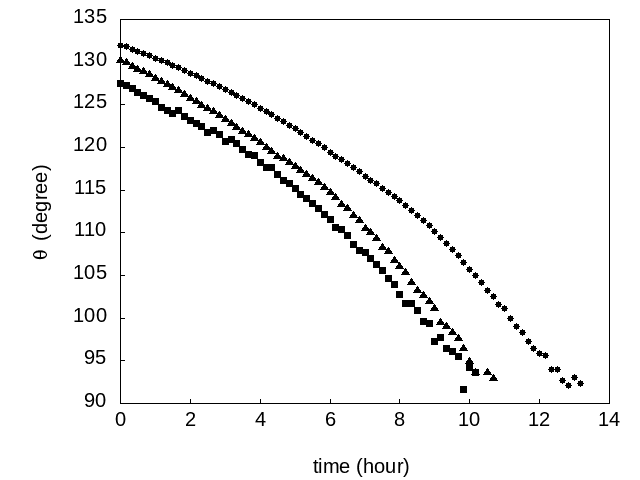}
    \caption{Contact angle of the droplets shown in Fig.S.\ref{fig:vt}. calculated from Suppl. eq.\ref{eq:supd}.}
    \label{fig:dt}
\end{figure}{}


\section{Droplet's radius of curvature as a function of time} 
\begin{figure}[H]
    \centering
    \includegraphics[width=8.6cm]{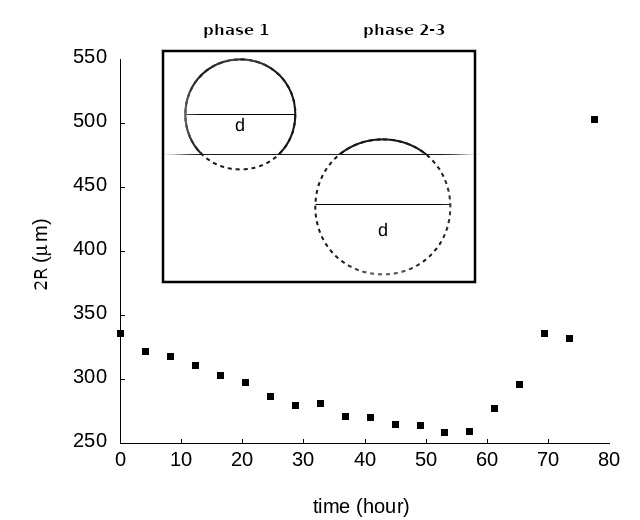}
    \caption{Droplet's radius of curvature determined on the basis of side view imaging. As it can be seen in the box, in phase 1 the radius decreases until the shape of the droplet reaches a hemisphere. After this minimum, the radius of curvature starts to increase in phases 2 and 3.}
    \label{fig:dt}
\end{figure}{}


\section{Microstructure formation with surfactant}

\begin{figure}[H]
    \centering
    \includegraphics[width=8.6cm]{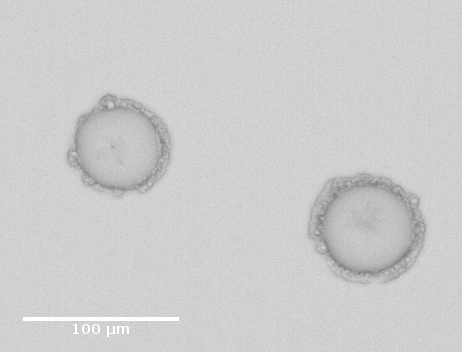}
    \caption{Water droplets in mineral oil supplemented with the surfactant Span80. After 2 hours the edge of the droplets started to exhibit a foam-like morphology, likely due to spontaneous microstructure formation.}
    \label{fig:emuks}
\end{figure}

\newpage
\section{Effect of capillary pressure}

In the theoretical analysis, the effect of capillary pressure on the saturation density of water is neglected due to the size of water molecules $R_\mathrm{w}\approx 1.4$~\AA{} being negligible compared to the droplet size $R$. In general, the saturation density of the dissolved water molecules in the vicinity of a curved oil-water interface increases with the curvature of the surface by a factor of $1+C_\mathrm{capillary}$, where the relative correction factor is given by [37]
\begin{equation}
 C_\mathrm{capillary} = \frac{2\sigma V_\mathrm{w}/R}{k_\mathrm{B}T} = \frac{2\sigma\frac{4}{3}{R_\mathrm{w}}^3\pi/R}{k_\mathrm{B}T},
\end{equation}
where $V_\mathrm{w}$ is the volume of a water molecule, $k_\mathrm{B}$ is the Boltzmann constant, $T$ is the absolute temperature, and $\sigma$ is the interfacial surface tension which originates from weak inter-molecular interactions and, therefore, can be expressed as:
\begin{equation}
    \sigma=\alpha\frac{k_\mathrm{B}T}{{R_\mathrm{w}}^2\pi}
\end{equation}
with a proportionality factor $\alpha$ being of the order of unity. Indeed, at around room temperature ($T\approx300$~K)
\begin{equation}
    \frac{k_\mathrm{B}T}{{R_\mathrm{w}}^2\pi}
    \approx 67\,\frac{\si{\milli\newton}}{\si\meter^2},
\end{equation}
which is comparable to the typical values of the oil-water surface tension, justifying that the coefficient $\alpha$ is of the order of unity. Thus, the capillary correction to the saturation density
\begin{equation}
    C_\mathrm{capillary}= \frac{8}{3}\alpha\cdot\frac{R_\mathrm{w}}{R}
\end{equation}
is negligible in our case, because $R_\mathrm{w}/R\ll1$.

\newpage
\section{Supplementary videos}
\renewcommand{\figurename}{Supplementary Video}
\setcounter{figure}{0}    
\begin{figure}[H]
    \centering
    \includegraphics[width=8.6cm]{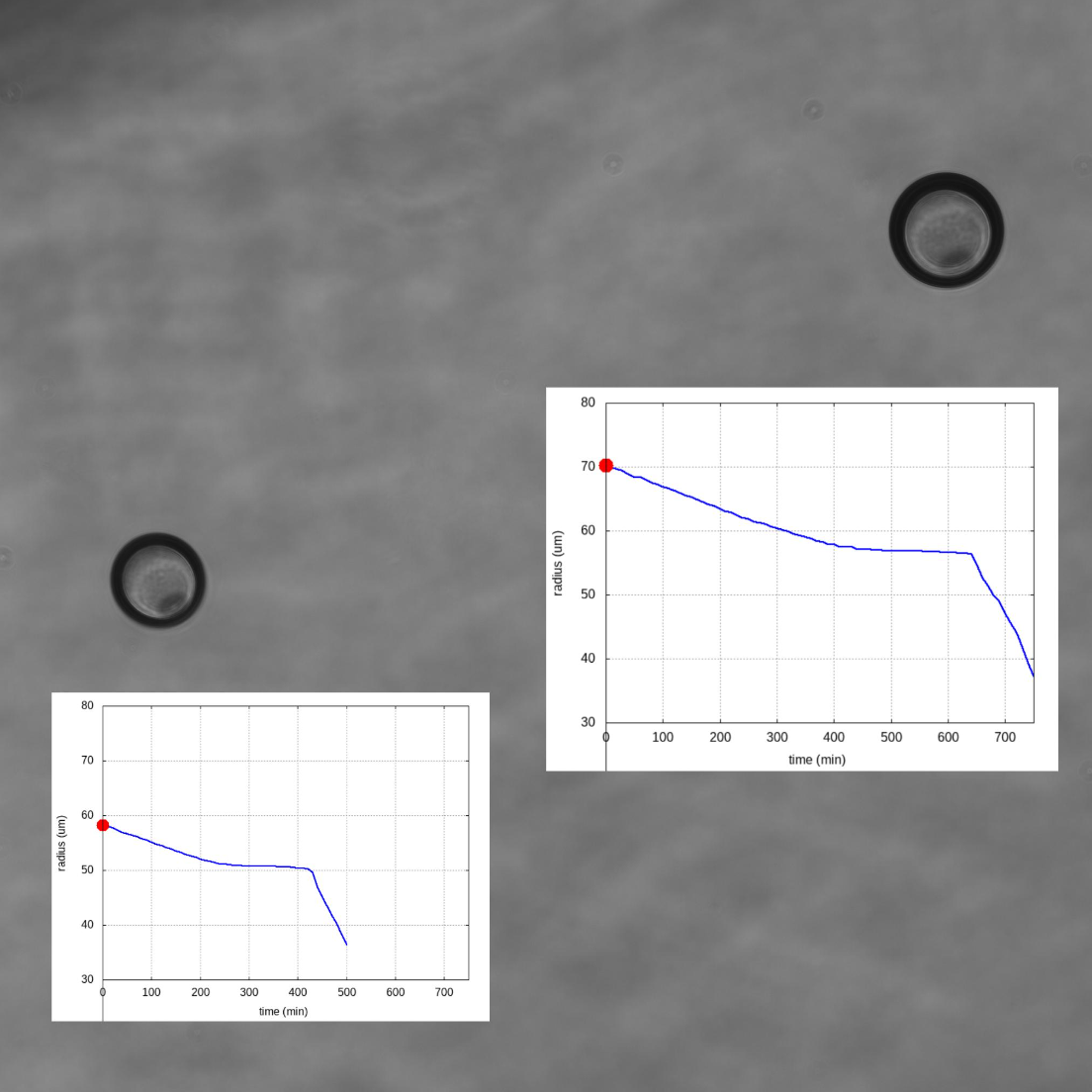}
    \caption{Time-lapse video showing the dissolution of two sessile droplets under mineral oil on a hydrophopic Petri dish surface imaged on an inverted microscope. The curves show the evolution of the radius of the corresponding droplets. From this bottom view the vertical projection of the droplet can be observed. The three phases of dissolution discussed in the paper can be well distinguished on the curves as well as on the images.}
    \label{fig:TLthumb}
\end{figure}{}

\begin{figure}[H]
    \centering
    \includegraphics[width=8.6cm]{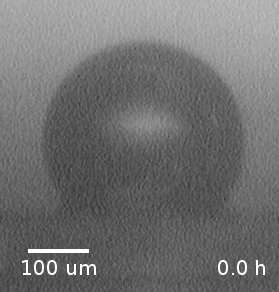}
    \caption{Dissolution of a sessile water droplet under mineral oil sitting on a hydrophobic plastic surface observed from a side view. Values of diameter and contact angle  as a function of time are shown in Fig.S5.. The droplet exhibits a spherical cap shape during the process, which takes 11.5 days to complete. }
    \label{fig:foblapse}
\end{figure}{}

\begin{figure}[H]
    \centering
    \includegraphics[width=8.6cm]{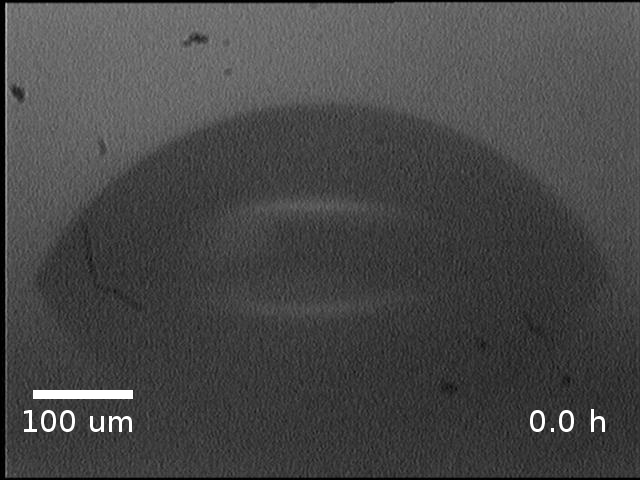}
    \caption{Dissolution of a sessile water droplet under mineral oil sitting on a hydrophilic glass surface observed from a side view. The complete dissolution of the droplet happens in a pinning mode, whereby the contact angle change compensates for volume loss while the contact area remains constant.}
    \label{fig:fillapse}
\end{figure}